\documentclass[11pt]{article}
\usepackage{cospar}
\usepackage{url}
\usepackage{xspace}
\usepackage{graphicx}
\usepackage{times}

\usepackage{natbib}
\usepackage[figuresright]{rotating}

\makeatletter

\renewcommand{\section}{\@startsection{section}{9}{\z@}{1\@bls plus
0.1\@bls}{1\@bls plus 0.1\@bls}{\normalsize\bf}}

\makeatother

\newcommand{\xmm}{{\it XMM-Newton}\xspace}
\newcommand{\chandra}{{\it Chandra}\xspace}
 
\newcommand\lax{\>\vcenter{\hbox{$<$\hskip-.75em\lower1.0ex\hbox{$\sim$}}}\>}
\newcommand\uax{\>\vcenter{\hbox{$>$\hskip-.75em\lower1.0ex\hbox{$\sim$}}}\>}
\title{XMM-NEWTON OBSERVATIONS OF THE VELA PULSAR}

\author{K. Mori$^1$, C.J. Hailey$^1$, F. Paerels
\address{Columbia Astrophysics
	Laboratory, 550 W 120th St., New York, NY 10027, USA} 
	and         
        S. Zane
\address{MSSL, University College London, Holmbury St. Mary,
 Dorking, Surrey, RH5 6NT, UK}}

\begin{document}

\maketitle

\begin{abstract}

We present spectral analysis from \xmm observations of the Vela
pulsar. We analyzed thermal emission from the pulsar dominating below $\sim 1$
keV since 
extracted spectra are heavily contaminated by nebular emission at higher
energy. Featureless high-resolution spectra of the Reflection Grating
Spectrometer aboard \xmm suggest the presence of a hydrogen atmosphere, as previously indicated by \chandra results. Both the 
temperature and radius are consistent with those values deduced from \chandra.
The derived \chandra and \xmm temperature of 
$T^{\infty}\simeq(6.4$--$7.1)\times10^5$ K at its age of $\sim10^4$ years is below
the standard  cooling curve.  

\end{abstract}

\section{INTRODUCTION} 

The Vela pulsar is one of the best-studied pulsars in
the radio band.  It has an 89 millisecond period and the spin-down parameters
indicate that the Vela pulsar is about $10^4$ years old with magnetic field
$3\times10^{12}$ G. Nevertheless,  multi-wavelength  observations in the past
two decades showed the Vela pulsar is a strong source at all wavelengths from
the radio to the gamma-ray band. In the X-ray band, thermal emission was
discovered by {\it ROSAT} \citep{ogelman93}.  {\it ROSAT} observation also
revealed a peculiar pulse profile with three peaks in the soft X-ray
band. Non-thermal emission from its surrounding nebula is dominant at higher
energies. 

Recent \chandra observations revealed even more complex properties of
the Vela pulsar and its nebula. With its high angular resolution, a \chandra image showed
detailed structure in  the surrounding nebula  \citep{helfand01,
pavlov01_2}. No spectral features were found  in the high energy resolution
\chandra LETGS spectra of the resolved pulsar emission. No detection of
spectral features excludes the presence of heavy element atmospheres on the
surface \citep{pavlov01}.  

 The Vela pulsar is also a strong source of non-thermal emission in the optical, X-ray and
gamma-ray bands. The non-thermal component from the pulsar was resolved for the
first time by \chandra \citep{pavlov01}. Recent {\it RXTE} observations
revealed complex timing properties associated with non-thermal emission from
the pulsar, showing energy-dependent multiple peaks in folded lightcurves
\citep{strickman99, harding02}. However, the origin of the peaks in the soft
X-ray band, whether thermal  or non-thermal, has yet to be determined.

The \xmm observation  of the Vela  pulsar was performed with a 90 ksec exposure time on
1 Dec 2000. \xmm is sensitive to faint sources such as isolated neutron stars
due to its large effective area. Since the nebula turned out to be compact
$\sim 10$--$20''$ \citep{helfand01}, the imaging capability of \xmm cannot
resolve the pulsar from the nebula. Therefore, the thermal component dominates
at $E  \lax  1$ keV and the nebula emission is dominant at $E \uax 1$ keV in \xmm
spectra. We could not spatially resolve the non-thermal emission from the pulsar or study different components of the nebula. Instead, using the \xmm data, we did
the following: (1) a spectral fit to high sensitivity Europian Photon Imaging
Camera (EPIC) \citep{struder01} data was used to
determine thermal properties with high accuracy, (2) a search for spectral
features  in high resolution Reflection Grating Spectrometer (RGS)
\citep{denherder01} data was performed.  All
the data were processed by version  5.3.3 of the Scientific Analysis System
(SAS) pipeline \citep{loiseau03}.   


\section{EPIC PHASE-AVERAGED SPECTRAL ANALYSIS}

We analyzed EPIC-PN data to determine the global features of the
X-ray spectrum. Source 
data were extracted from a circular region with radius $r=1'$ and events with
CCD pattern 0 were selected. Such a source area includes more than 90\% of the
source counts at energies up to $\sim$ 5 keV. However, the extracted spectra
are heavily contaminated by nebular emission at $E \uax 1$
keV. Unlike the recent \chandra results which resolved the pulsar emission
from its nebula emission 
\citep{pavlov01}, our analysis focuses on thermal emission from the pulsar at
$E \lax 1$ keV. Systematic errors from nebular contamination are discussed
later in this section.  Background  spectra were reduced
from an annulus region between $r=1'$ and $r=2'$. Total EPIC-PN count rate from
the $r<1'$ region after subtracting background was  27.0 cts/sec.  

\begin{figure}[!t]
\centering 
\includegraphics[width=120mm]{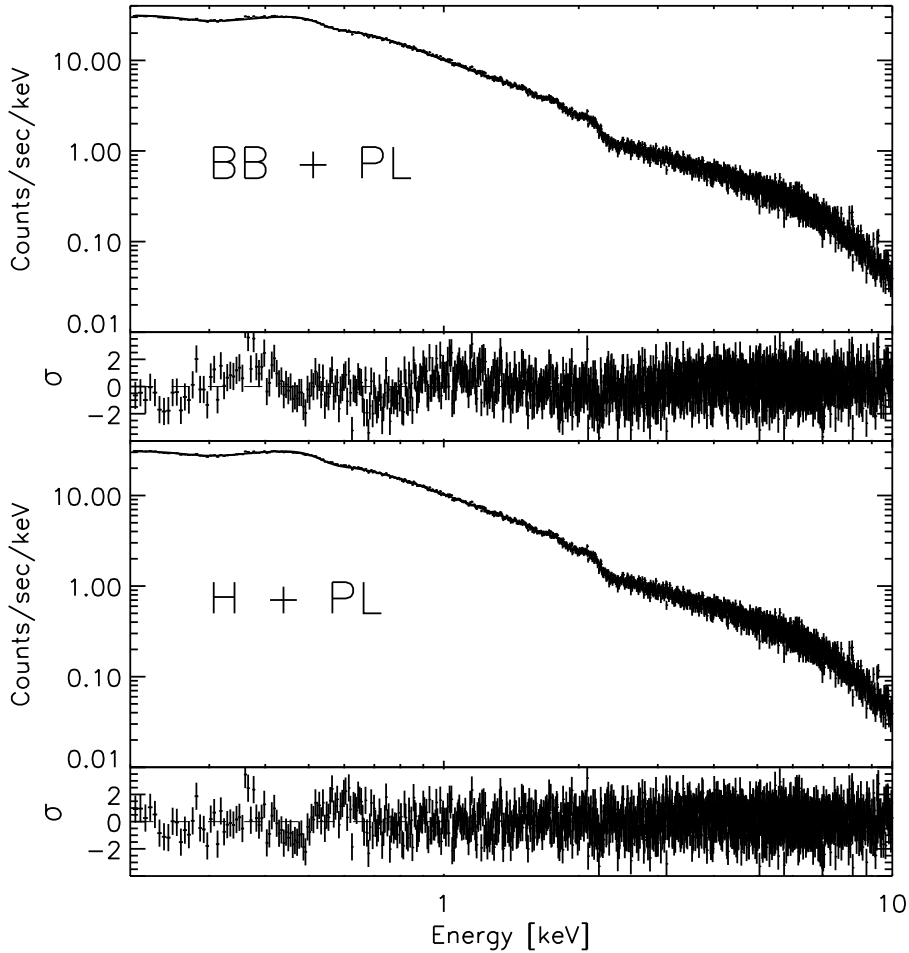}
\vskip -8mm

\caption{\ \xmm/EPIC-PN spectra fitted by Blackbody+Power-law (BB+PL) (top)
and Hydrogen atmosphere+Power-law model (H+PL) (bottom). Residuals at $\sim0.5$ keV are due to instrumental Oxygen
edge. \label{fig_pn}}
\end{figure} 

We rebinned 
the data so that each spectral bin contained more than 25 counts. For the
following spectral analysis, we used ready-made Response Matrix Files (RMFs)
and Auxiliary Response Files (ARFs) generated by
the SAS tool ``arfgen''. For spectral fitting, we adopted a two-component model
comprised of a thermal and non-thermal (power-law) spectrum. We applied two
different thermal models to the data: a blackbody model (BB+PL model) and a
magnetized hydrogen atmosphere model provided by V. Zavlin \citep{zavlin03} at $B=10^{12}$G (H+PL model). When we
fitted a hydrogen atmosphere model, we fixed the neutron star mass and radius
to $1.4M_{\odot}$ and 10 km. Both thermal models fit the data well, yielding
$\chi^2\simeq1.1$ (Figure 1  and Table 1). Figure 2 shows an $N_H$--$T^{\infty}$
contour plot with fixed power-law index and radius. Hydrogen atmospheres are in
general harder than a blackbody since free-free absorption dominates in the
X-ray band. Therefore, the temperature fitted by a magnetized
hydrogen atmosphere model is lower than a blackbody fit (hence a larger radius
for the hydrogen atmosphere fit).    

The derived distance ($d=256$ pc) is consistent with the distance from the
parallax measurement ($d=294^{+76}_{-50}$  
pc) \citep{caraveo01}. For hydrogen atmosphere models,  $T^{\infty}$ and
$R^{\infty}/d$ depend on other input parameters (e.g. $M$ and $R$) weakly  \citep{zavlin98}. The
apparent radius corresponding to the measured distance is inferred to be
$R^{\infty}/d_{294}=15.0_{-5.1}^{+6.5}$ km where $d_{294}$ is the distance in
units of 294 pc.      

\begin{figure}[!t]
\centering 
\includegraphics[width=120mm]{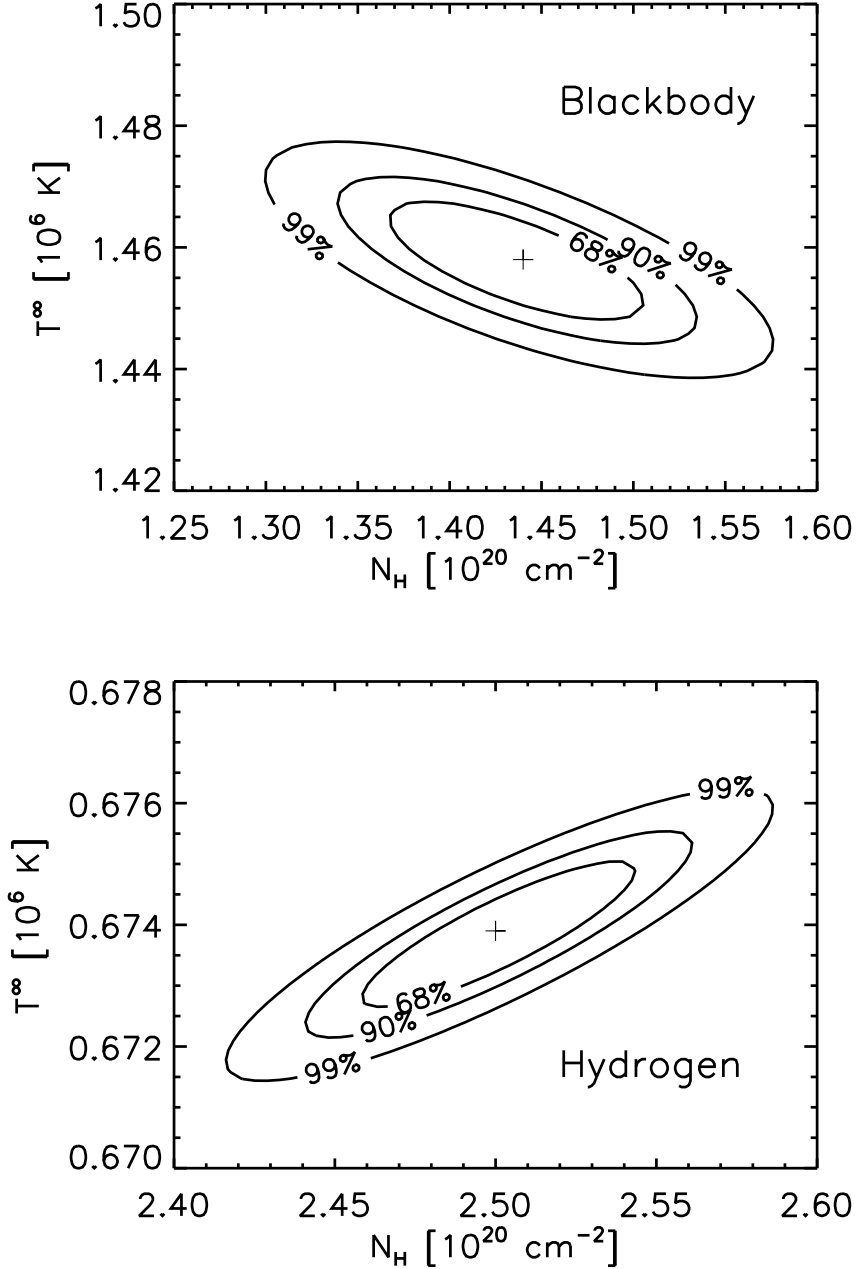}
\vskip -8mm 

\caption{\ Contour plot in $N_H$--$T^{\infty}$ plane. Blackbody (top)
and hydrogen atmosphere at $B=10^{12}$ G (bottom). \label{fig_cont}} 
\end{figure}


\begin{table}[!t]
\centering 
\vspace{-8mm}
\begin{minipage}{92mm}  
\caption{\ Fitted parameters to \xmm EPIC-PN spectra (pulsar plus nebula) in
comparison with \chandra data (pulsar only).
\label{tab_fit}}
\centerline{Blackbody + Power-law}
\begin{tabular}{lcc}
\hline
& EPIC-PN     &  \chandra  \\ 
\hline
$N_H$ [$10^{20}$ cm$^{-2}$]     & $1.44^{+0.83}_{-0.85}$ & $ 2.2\pm0.3 $
\\
$T^{\infty}$ [$10^6$ K] & $1.46_{-0.02}^{+0.02}$ &  $1.50\pm0.05$   \\
$R^{\infty}$ [km]$^{\mbox{a}}$ & $2.50_{-0.14}^{+0.15}$ & $ 2.5\pm0.2 $   \\
$L^{\infty}_{th}$ [$10^{32}$ ergs\,s$^{-1}$]$^{\mbox{a}}$ & 	$2.02_{-0.33}^{+0.34}$ & $2.1\pm0.3$\\
$\gamma$ $^{\mbox{b}}$ 	& $1.64_{-0.08}^{+0.08}$ & $ 2.7\pm0.4 $ \\
$L_{pl}$ [$10^{32}$ ergs\,s$^{-1}$]$^{\mbox{a},\mbox{b}}$ & $8.56_{-0.14}^{+0.11} $
&$0.58\pm0.08$	\\
$\chi^2_\nu$    		& $1.16$    	&    $1.1$ \\
\hline
\end{tabular}
\vspace{0.4cm}

\centerline{Magnetized H atmosphere $^c$ + Power-law}
\begin{tabular}{lcc}
\hline        
& EPIC-PN   &  \chandra  \\ 
\hline
$N_H$ [$10^{20}$ cm$^{-2}$]  & $2.50_{-0.17}^{+0.16}$    &   $3.3\pm0.3$ \\
$T^{\infty}$ [$10^6$ K] &  $0.674_{-0.033}^{+0.034}$	&   $0.68\pm0.03$\\
$d$ [pc]  & $ 256_{-43}^{+44}$  &   $210\pm20$	 \\
$L^{\infty}_{th}$ [$10^{32}$ ergs\,s$^{-1}$]$^{\mbox{a}}$ &
$3.35_{-1.77}^{+1.78} $ &$5.2\pm0.4$	\\
$\gamma$ $^{\mbox{b}}$& $1.59_{-0.12}^{+0.12}$    & $1.5\pm0.3$     \\
$L_{pl}$ [$10^{32}$ ergs\,s$^{-1}$]$^{\mbox{a},\mbox{b}}$	& $8.27_{-0.14}^{+0.14}$	& $0.21\pm0.06$	\\
$\chi^2_\nu$    		& $1.08$            &$1.0$ \\
\hline
\end{tabular}
\vspace{0.2cm}

{Errors for EPIC-PN data are 1 sigma level including both statistical and
systematic uncertainty. For luminosity of non-thermal component in
 the range of 2--10 keV ($L_{pl}$), the 
error refers only to the statistical uncertainty since the smaller extracted region
picks up less nebular emission.} \\
$^{\mbox{a}}$ We assumed the distance is 294 pc from the latest parallax
measurement \citep{caraveo01}. We rescaled the \chandra results in Table 1 in \citet{pavlov01} to $d=294$
pc. \\ 
$^{\mbox{b}}$ Note that these parameters are
different from the \chandra results in comparison. However, the fitted
parameters to the \xmm data is dominated by the
nebular component, while \citet{pavlov01} extracted spectra from the pulsar.  
\\
$^{\mbox{c}}$ We fixed the neutron star mass to $1.4M_{\odot}$ and
the radius to 10 km for spectral fitting. Here $B=10^{12}$ G.  
\end{minipage}
\end{table} 

\begin{figure}[!t] 
\centering 
\includegraphics[width=115mm]{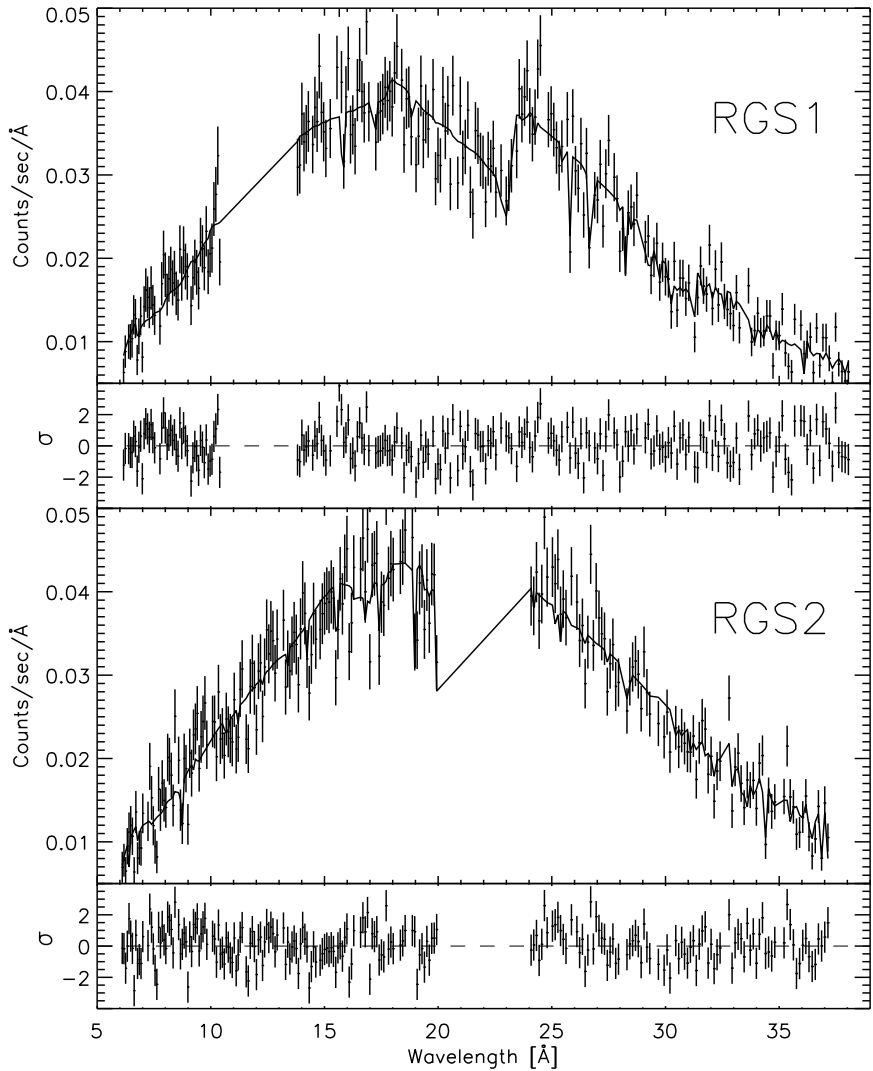}

\caption{\ \xmm/RGS spectra fit by Hydrogen+Power-law  
model. The spectra were extracted from the 70\% Point Spread Function (PSF) region. \label{fig_rgs}}
\end{figure} 

We note that non-thermal spectra from the pulsar (obtained by
\chandra observation) are about two orders of magnitude weaker than the
nebular spectra, so the power-law component observed by \xmm is dominated by
the nebular contribution. The fitted power-law index ($\gamma\simeq1.6$)
probably represents a superposition of different nebular components, which were
resolved by the \chandra observation \citep{pavlov01_2}. The estimated
luminosity of the power-law component in the EPIC-PN data is consistent with
the \chandra result for the nebula ($L_{neb}=8.3\times10^{32} $ ergs\,s$^{-1}$
for $d=294$ pc) \citep{pavlov01_2}.    

The derived parameters from both the blackbody and magnetized hydrogen atmosphere fit are
in good agreement with  the \chandra results. 
However, since the angular resolution of \xmm is poorer than that of
\chandra, our fitting parameters can be affected by systematic errors due the
the nebular contamination. These errors are, in
general, larger than the statistical ones.
In order to estimate such errors, we examined the variations in the hydrogen atmosphere and
blackbody parameters by using as extraction regions annuli of various
radii; the errors reported in Table 1 reflect both statistical and
systematic uncertainties. We can see that, when considering the effect on
the nebular contamination, \chandra and \xmm temperatures are in
agreement at 1-sigma level.


\section{SEARCH FOR SPECTRAL FEATURES IN RGS DATA}

Reflection Grating Spectrometer data were taken in ``spectroscopy'' mode in order to take full advantage of 
 the high spectral resolution. First, we extracted the 95\% Point Spread
 Function (PSF) of the source
 region for both RGS 1 and 2 data.  The background spectrum was reduced from outside the 98\% PSF
 region. We generated responses by the SAS command  ``rgsrmfgen''.  Total
 count rate was 1.0 cts/sec for both RGS 1 and 2 data.  We extracted smaller
 regions of the source (50\% and 70\% PSF region) to maximize the number of
 thermal photons compared to nebular emission photons. Figure 3 shows RGS
 spectra extracted from the 70\% PSF region and fitted by the H+PL model.  The
 data are plotted in the RGS bandpass ($\lambda = 6$--$38$ \AA) after
 rebinning wavelength grids to $\lambda\sim0.1$\AA.  

The spectral fit by both
 the BB+PL and the H+PL model yield $\chi^2_{\nu}\simeq1.2$ for both RGS 1 and
 2 data. The fitted parameters were similar to the EPIC-PN results in Table 1.
 Several  deviations from the continuum fit with less than 2 sigma significance
 are seen, but none of these wiggles were found in both RGS 1 and 2 data
 simultaneously. Therefore, we conclude that the RGS spectra are featureless.      


\section{SUMMARY} 



The results obtained with \xmm for the Vela pulsar are in agreement
with previous findings by \chandra \citep{pavlov01, pavlov01_2}. The RGS data indicate that the thermal spectrum is featureless,
and the EPIC-PN spectrum is well fitted by a magnetized hydrogen
atmosphere model with $B=10^{12}$ G. As it has been found with \chandra,
with respect to a fit with the simple blackbody, 
the atmospheric model has the advantage that it gives a radius for the
emitting region compatible with neutron star equations of state. Also,
when \chandra data are fitted with an atmospheric model, the index of
the power-law component above $\sim 1$ keV is remarkably consistent
with that of the PL observed at optical and hard X-ray wavelengths. The
power-law component observed with \xmm is dominated by the nebular
emission and the value of $L_{pl} \sim 8\times10^{32}$ erg\,s$^{-1}$ is again consistent
with that observed by \chandra \citep{pavlov01_2}. 

The presence of hydrogen is not surprising since a tiny amount of hydrogen
constitutes an optically-thick layer in the atmosphere \citep{zavlin02}. Hydrogen could
be accumulated on the surface by accretion or generated by spallation of
fallback material \citep{bildsten92}. The derived large radius from the H+PL
model implies that the  thermal emission originates from the whole
surface.  

There is an important implication for the neutron star physics
which is consistent with the \chandra and \xmm results. Surface  temperature for a given age reflects cooling processes via neutrino emission in the core. Recent studies show that
neutron star cooling curves are dependent on various factors such as
proton-to-neutron ratio, neutron star mass, magnetic field strength, neutron
superfluidity and presence of exotic matter. Among various proposed models, a
model which assumes normal composition ($n,p$ and $e$) and the indirect URCA
process was often quoted as the standard cooling model. The standard cooling
model fits most cooling neutron stars, although recent X-ray observations
found several neutron stars have surface temperatures below the
standard cooling curve \citep{slane02}. For the Vela pulsar, the derived
\chandra and \xmm temperature of $T^{\infty}\simeq(6.4$--$7.1)\times10^5$ K at its
age $\sim10^4$ years is below the standard  cooling curve \citep{tsuruta02}. 
   
\section*{ACKNOWLEDGEMENTS}

We are grateful to V. Zavlin for providing us with the magnetized hydrogen atmosphere
model.




\begin{thebibliography}{}

\bibitem[{{Bildsten} {et~al.}(1992)}]{bildsten92} 

{Bildsten}, L. and {Salpeter}, E.~E. \&  {Wasserman}, I., The fate of
accreted CNO elements in neutron star atmospheres - X-ray bursts and gamma-ray
lines, {\em Astrophys. J.}, {\bf 384}, 143-176, 1992. 

\bibitem[{{Caraveo} {et~al.}(2001)}]{caraveo01}

{Caraveo}, P.~A., {De Luca}, A., {Mignani}, R.~P. et al.,
The distance to the Vela pulsar gauged with Hubble Space Telescope
parallax observations, {\em Astrophys. J.}, {\bf 561}, 930-937, 2001. 

\bibitem[{{den Herder} {et~al.}(2001)}]{denherder01} 

{den Herder}, J.~W., {Brinkman}, A.~C., {Kahn}, S.~M. et al, The Reflection
Grating Spectrometer on board XMM-Newton, {\em Astron. Astrophys.}, 365, L7-L17,
2001.  

\bibitem[{{Harding} {et~al.}(2002)}]{harding02}

{Harding}, A.~K., {Strickman}, M.~S., {Gwinn}, C. et al., The multicomponent nature of the Vela pulsar
non-thermal X-Ray spectrum,  {\em Astrophys. J.}, {\bf 576}, 376-380, 2002.  

\bibitem[{{Helfand} {et~al.}(2001)}]{helfand01}

{Helfand}, D.~J., {Gotthelf},
E.~V., \& {Halpern}, J.~P., Vela pulsar and its synchrotron nebula, {\em
Astrophys. J.}, {\bf 556}, 380-391, 2001. 

\bibitem[{{Loiseau} (2003)}]{loiseau03} 

{Loiseau}, N., Users' Guide to the XMM-Newton Science Analysis System, 
http://xmm.vilspa.esa.es/external/
xmm\_user\_support/documentation/index.shtml,  2003. 

\bibitem[{{{\"O}gelman} {et~al.}(1993)}]{ogelman93}

{{\"O}gelman}, H., {Finley}, J.~P., \& {Zimmerman}, H.~U., Pulsed X-rays
from the Vela pulsar, {\em Nature}, {\bf 361}, 136-138, 1993. 

\bibitem[{{Pavlov} {et~al.}(2001a)}]{pavlov01} 

{Pavlov}, G.~G., {Zavlin},
V.~E., {Sanwal}, D. et al.,  The X-ray spectrum of
the Vela pulsar resolved with Chandra, {\em Astrophys. J.}, {\bf 552},
L129-L133, 2001a.  

\bibitem[{{Pavlov} {et~al.}(2001b)}]{pavlov01_2}

{Pavlov}, G.~G., {Kargaltsev}, O.~Y., {Sanwal}, D. et al., Variability of the Vela pulsar wind nebula observed with
Chandra, {\em Astrophys. J.}, {\bf 554}, L189-L192, 2001b.  

\bibitem[{{Slane} {et~al.}(2002)}]{slane02}

{Slane}, P.~O., {Helfand}, D.~J.,
\& {Murray}, S.~S., New constraints on neutron star cooling from Chandra
observations of 3C 58 {\em Astrophys. J.}, {\bf 571}, L45-L49, 2002. 

\bibitem[{{Strickman} {et~al.}(1999)}]{strickman99}

{Strickman}, M.~S.,
{Harding}, A.~K., \& {de Jager}, O.~C., A Rossi X-Ray timing explorer
observation of the Vela pulsar: filling in the X-Ray gap, {\em Astrophys. J.},
{\bf 524}, 373-378, 1999.    

\bibitem[{{Str{\"u}der} {et~al.}(2001)}]{struder01} 
{Str{\"u}der}, L., {Briel}, U., {Dennerl}, K. et al., The European Photon
Imaging Camera on XMM-Newton: The pn-CCD camera {\em Astron. Astrophys.}, L18-L26,
2001.  

\bibitem[{{Tsuruta} {et~al.}(2002)}]{tsuruta02}

 {Tsuruta}, S., {Teter}, M.~A.,
{Takatsuka}, T. et al., Confronting neutron star
cooling theories with new observations, {\em  Astrophys. J.}, {\bf 571}, L143-L146,
2002.   

\bibitem[{{Zavlin} {et~al.}(1998)}]{zavlin98}

{Zavlin}, V.~E., {Pavlov}, G.~G., \& {Trumper}, J., The neutron star in the
supernova remnant PKS 1209-52, {\em Astron. Astrophys.}, 331, 821-828, 1998.  

\bibitem[{{Zavlin} \& {Pavlov}(2002)}]{zavlin02}

{Zavlin}, V.~E. \& {Pavlov}, G.~G., Modeling neutron star atmospheres, {\em
proceedings of the 270. WE-Heraeus Seminar on Neutron Stars, Pulsars and
Supernova Remnants}, MPE-Report 278 (astro-ph/0206025), 263-272, 2002.  

\bibitem[{{Zavlin} (2003)}]{zavlin03} 

{Zavlin}, V.~E., private communication, 2003. \\

\end{thebibliography}
\end{document}